\documentclass[conference]{IEEEtran}
\IEEEoverridecommandlockouts

\usepackage{cite}
\usepackage{amsmath,amssymb,amsfonts}
\usepackage{graphicx}
\usepackage{textcomp}
\usepackage{xcolor}
\usepackage{subcaption}

\def\BibTeX{{\rm B\kern-.05em{\sc i\kern-.025em b}\kern-.08em
    T\kern-.1667em\lower.7ex\hbox{E}\kern-.125emX}}

\begin{document}

\title{AI-Enhanced Virtual Biopsies for Brain Tumor Diagnosis in Low Resource Settings}

\author{\IEEEauthorblockN{Areeb Ehsan}
\IEEEauthorblockA{\textit{Department of Computer Science} \\
\textit{Georgia State University}\\
Atlanta, GA, USA \\
aehsan1@student.gsu.edu}
}

\maketitle

\begin{abstract}
Timely brain tumor diagnosis remains challenging in low-resource clinical environments where expert neuro-radiology interpretation, high-end MRI hardware, and invasive biopsy procedures may be limited. Although deep learning has achieved strong performance in brain tumor analysis, real-world adoption is still limited by major issues like computational demands, dataset shift across scanners, and limited interpretability. This paper presents a prototype ``virtual biopsy'' pipeline that can predict brain tumor class from 2D MRI images using a lightweight convolutional neural network (CNN) and complementary radiomics-style features. A MobileNetV2-based CNN is trained for four-class classification, while an interpretable radiomics branch extracts eight handcrafted features capturing lesion shape, intensity statistics, and gray-level occurrence matrix (GLCM) texture descriptors. A hybrid fusion strategy concatenates CNN latent embeddings with radiomics features and trains a RandomForest classifier on the fused representation. Explainability is provided through Grad-CAM heatmaps along with radiomics feature importance analysis. Performing experiments on a public Kaggle brain tumor MRI dataset demonstrates that the fusion improves validation performance relative to single-branch baselines, while robustness tests under low-resolution and noisy conditions highlight sensitivity of CNN-based methods which is relevant to low-resource environments facing limitations. Furthermore, discussions have been done on framing the system as a decision support rather than a full-on diagnostic replacement.
\end{abstract}


\section{Introduction}
Brain and central nervous system tumors pose a substantial clinical challenge, with outcomes strongly dependent on timely diagnosis and appropriate treatment planning \cite{b12,b13}. In many settings, magnetic resonance imaging (MRI) is the primary non-invasive gold-standard that is used to identify suspicious lesions and guide any interventions. However, these diagnostic workflows often rely on expert radiologist interpretation and when feasible, surgical biopsy followed by histopathology to confirm tumor type and grade. Biopsy is invasive and may be contraindicated for tumors located in surgically sensitive regions which may make the procedure risky, in addition, biopsy resources and pathology infrastructure may be limited in rural or underfunded hospitals or clinics. These constraints motivate research into a set of``virtual biopsy'' methods which can estimate clinically relevant tumor characteristics directly from imaging in order to support earlier and safer decision-making.

Deep learning has demonstrated strong performance for brain tumor segmentation and classification. Nevertheless, three recurring barriers impede deployment in low-resource environments: (i) computational requirements of state-of-the-art 3D CNNs and large models, (ii) limited generalization due to dataset shift across scanners and acquisition protocols \cite{b6,b12}, and (iii) ``black-box'' decision-making that can reduce clinician trust. In response, this work develops a prototype pipeline that combines a lightweight CNN backbone with radiomics-style interpretable descriptors and provides explainability via Grad-CAM. The aim is to explore a practical, CPU-feasible design while explicitly characterizing robustness under degraded or low-resource imaging conditions which can occur in resource constrained environments.

\section{Related Work}
U-Net and its variants are widely used for MRI tumor segmentation, enabling pixel-level delineation of tumor regions that can support downstream analysis \cite{b1,b6,b8,b9}. CNN-based classifiers have been applied to tumor detection and type classification [7], including VGG-derived architectures \cite{b2} and efficiency-oriented backbones designed for lower compute budgets. Beyond centralized training, approaches such as federated learning have been explored to address privacy and data fragmentation challenges and to expand training data diversity in oncology imaging \cite{b3}.

Radiomics extracts handcrafted features from medical images to quantify lesion shape, intensity distribution, and texture, providing interpretable descriptors that may complement deep features. Hybrid strategies that combine radiomics with deep representations are increasingly studied to balance performance with interpretability and robustness. Recent work has also emphasized explainability frameworks for brain tumor imaging, including neuro-fuzzy or rule-based reasoning to translate features into clinician-friendly explanations \cite{b5}. Despite progress, practical adoption remains challenging in settings where imaging quality and hardware vary substantially, underscoring the need for prototype evaluations that explicitly probe robustness.

\section{Dataset and Preprocessing}
\subsection{Dataset}
Prototype-level experiments were conducted using a publicly available brain tumor MRI image classification \cite{b10,b11} dataset organized in a directory structure suitable for \texttt{ImageFolder} loading (Kaggle: ``Brain Tumor Classification (MRI)'' by Sartaj Bhuvaji). The dataset provides four classes: glioma tumor, meningioma tumor, pituitary tumor, and no tumor. The training split contains approximately 2{,}870 images, which was further partitioned into an 80/20 train/validation split, yielding 2{,}296 training images and 574 validation images in the reported setup. A separate test split contains 394 images.

The dataset is comprised of 2D MRI images; specific MRI sequence metadata (e.g., T1, T2, FLAIR) is not explicitly provided in the dataset used here. Accordingly, the analysis treats the input as a single-sequence 2D MRI image classification task and avoids sequence-specific claims.
\begin{figure}[t]
\centering
\includegraphics[width=\linewidth]{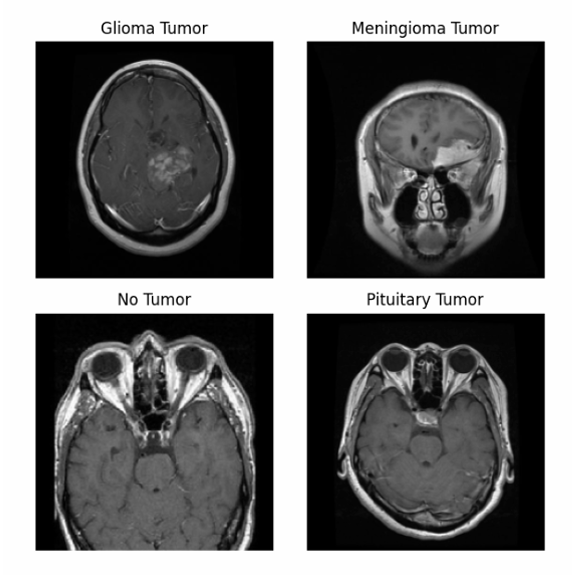}
\caption{Representative sample MRI images from the Kaggle dataset across the four classes: glioma tumor, meningioma tumor, pituitary tumor, and no tumor.}
\label{fig:dataset}
\end{figure}

\subsection{Preprocessing}
All images were resized to $224 \times 224$ pixels and normalized to match the requirements of ImageNet-pretrained CNN backbones. The applied transform pipeline consisted of resizing, conversion to tensor format, and per-channel normalization using mean $0.5$ and standard deviation $0.5$ (mapping input intensities to approximately $[-1,1]$). DataLoaders were constructed with a batch size of 16. Training batches were shuffled, while validation and test batches were not shuffled.

\section{Proposed Method}
The proposed ``virtual biopsy'' prototype integrates (i) a lightweight CNN classifier, (ii) radiomics-style feature extraction, (iii) a hybrid feature fusion classifier, and (iv) explainability via Grad-CAM and feature importance.

\begin{figure}[t]
\centering
\includegraphics[width=0.95\linewidth]{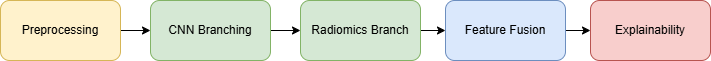}
\caption{Overall prototype pipeline for CNN--radiomics fusion-based ``virtual biopsy'' classification and explainability.}
\label{fig:system}
\end{figure}

\subsection{CNN Architecture}
\label{sec:cnn}
A MobileNetV2 backbone pretrained on ImageNet was used as the CNN branch due to its computational efficiency and suitability for CPU inference. The final classification layer was replaced to output logits for four classes. Let $\mathbf{x} \in \mathbb{R}^{224 \times 224 \times 3}$ denote the normalized input image. The CNN produces class logits
\begin{equation}
\mathbf{z} = f_{\theta}(\mathbf{x}) \in \mathbb{R}^{4},
\end{equation}
where $f_{\theta}$ denotes the MobileNetV2 network with parameters $\theta$. Cross-entropy loss was used for training, and the model was optimized with Adam at learning rate $10^{-4}$ for three epochs on CPU in the reported prototype.

In addition to logits, the penultimate CNN representation (the MobileNetV2 latent embedding) was used for fusion. MobileNetV2 produces a 1280-dimensional feature vector per image after global pooling; this embedding is denoted by $\mathbf{h} \in \mathbb{R}^{1280}$.

\begin{figure}[t]
\centering
\includegraphics[height=5.2cm,keepaspectratio]{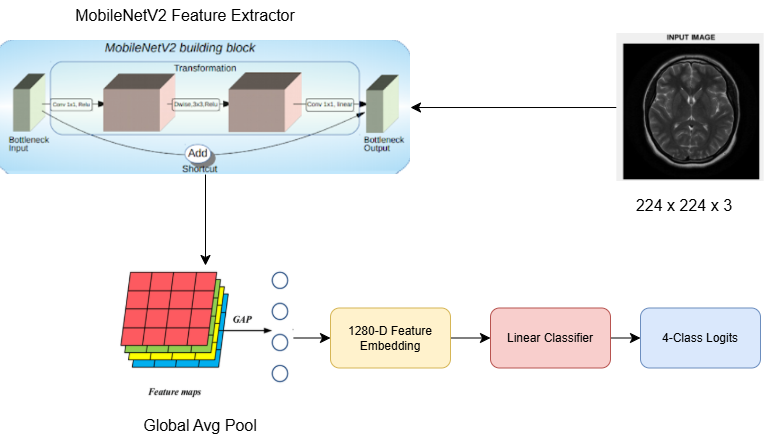}
\caption{CNN branch using MobileNetV2 for four-class tumor classification and for extracting 1280-dimensional embeddings for fusion.}
\label{fig:cnn}
\end{figure}

\subsection{Radiomics Feature Extraction}
\label{sec:radiomics}
A radiomics-style branch was implemented to compute interpretable features from each image. The image tensor was first converted back to grayscale intensity in $[0,1]$ by inverting normalization and averaging channels when needed [8]. A simple tumor mask was then obtained using Otsu thresholding on the grayscale image. Connected component analysis was used to identify the largest region, from which basic region properties were computed.

Eight features were extracted per image and grouped as follows.

\textit{Shape features:} area, eccentricity, and solidity computed from the largest connected component of the binary mask.

\textit{Intensity features:} mean intensity and standard deviation of intensity within the region (or globally if region extraction fails).

\textit{Texture features:} GLCM-based contrast, homogeneity, and entropy computed on an 8-bit quantized image. 

This branch is intended as an interpretable approximation rather than a full-fledge clinically approved radiomics pipeline. In particular, Otsu-based masking is a lightweight heuristic that may fail on low-contrast or heterogeneous scans.

\begin{figure}[t]
\centering
\includegraphics[width=\linewidth]{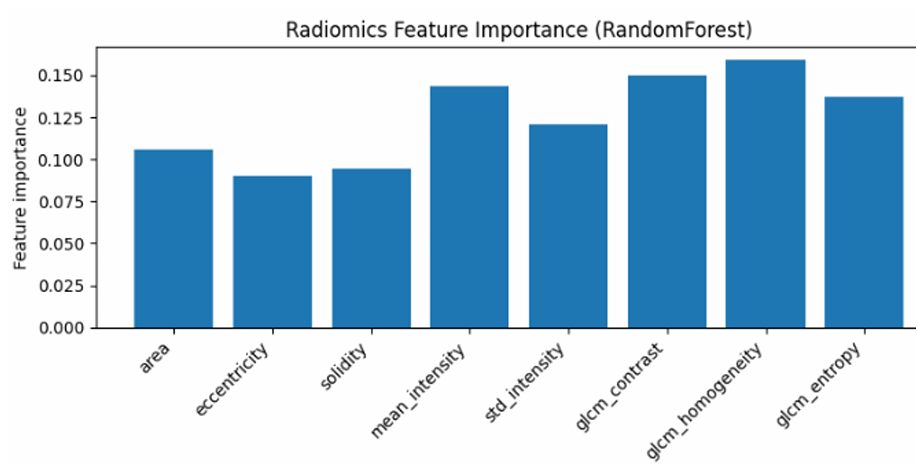}
\caption{Radiomics feature importance from the RandomForest model, providing interpretable cues about which handcrafted descriptors contributed most to predictions.}
\label{fig:radiomics}
\end{figure}

\subsection{Feature Fusion Strategy}
\label{sec:fusion}
A hybrid (late) fusion strategy was used [9]. CNN embeddings $\mathbf{h} \in \mathbb{R}^{1280}$ were concatenated with radiomics features $\mathbf{r} \in \mathbb{R}^{8}$ to form a fused feature vector
\begin{equation}
\mathbf{u} = [\mathbf{h} \, \Vert \, \mathbf{r}] \in \mathbb{R}^{1288},
\end{equation}
where $\Vert$ denotes vector concatenation. A RandomForest classifier was trained on $\mathbf{u}$ so that it can predict the class label. This was chosen since the ability of tree ensembles to handle mixed feature types and provide feature importance measures that support interpretability.

\begin{figure}[t]
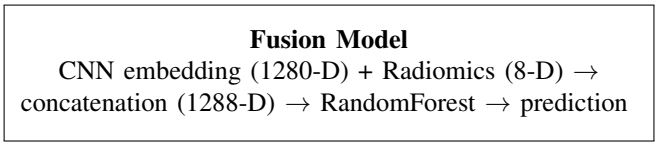

\centering
\fbox{\parbox{0.95\linewidth}{\centering \vspace{6pt}
\textbf{Fusion Model}\\
CNN embedding (1280-D) + Radiomics (8-D) $\rightarrow$ concatenation (1288-D) $\rightarrow$ RandomForest $\rightarrow$ prediction
\vspace{6pt}}}
\caption{Hybrid fusion strategy concatenating deep embeddings and radiomics features, followed by a RandomForest classifier.}
\label{fig:fusion}
\end{figure}

\subsection{Explainability (Grad-CAM)}
\label{sec:gradcam}
Grad-CAM was also implemented for the CNN branch to highlight regions that most influenced predictions [12]. Grad-CAM computes a coarse localization map by weighting the convolutional feature maps by spatially pooled gradients of the target class score with respect to those maps. Let $A^k$ denote the $k$-th feature map in the selected convolutional layer, and let $y^c$ denote the score for class $c$. Grad-CAM weights are computed as
\begin{equation}
\alpha_k^c = \frac{1}{Z}\sum_{i}\sum_{j}\frac{\partial y^c}{\partial A^k_{ij}},
\end{equation}
where $Z$ is the number of spatial locations. The Grad-CAM heatmap is then
\begin{equation}
L_{\mathrm{GradCAM}}^c = \mathrm{ReLU}\left(\sum_{k}\alpha_k^c A^k\right).
\end{equation}
In the prototype, hooks were registered on the last convolutional block of MobileNetV2 to capture activations and gradients. The resulting heatmap was resized to the input resolution and overlaid on the original image for visualization.

Additional explainability was provided by reporting radiomics feature importances from the RandomForest model, which enabled numerical explanation alongside the visual heatmaps.

\begin{figure}[t]
\centering
\includegraphics[width=\linewidth]{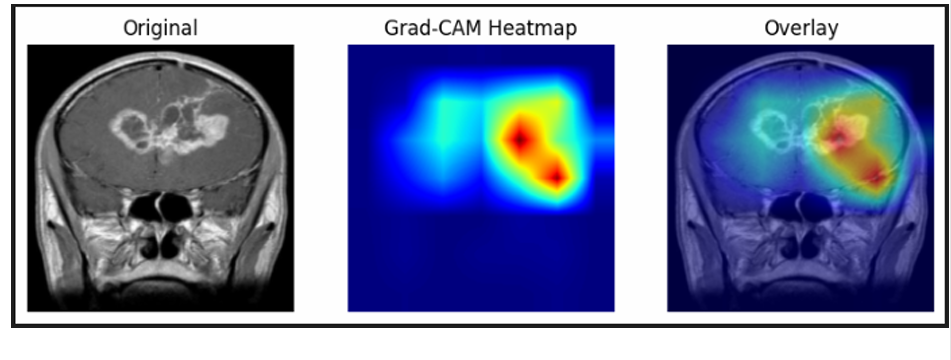}
\caption{Grad-CAM visualization for CNN predictions. Heatmaps indicate salient regions contributing to the predicted class.}
\label{fig:gradcam}
\end{figure}

\section{Experimental Setup}
Training and evaluation were performed in a Jupyter notebook prototype implementation using PyTorch for CNN training and Scikit-Learn for classical models. The CNN baseline was trained for three epochs on CPU with Adam optimizer and cross-entropy loss. Model evaluation used accuracy and macro-averaged F1 score to reduce sensitivity to class imbalance. For a dataset with $C$ classes, macro-F1 is defined as
\begin{equation}
\mathrm{F1_{macro}} = \frac{1}{C}\sum_{c=1}^{C}\mathrm{F1}_c.
\end{equation}
Radiomics-only and fusion models used RandomForest classifiers (with preprocessing via a Scikit-Learn pipeline where applicable). For fusion, CNN embeddings were extracted for each image and concatenated with the radiomics features.

To simulate low-resource conditions, two combination of families were evaluated on the following CNN baseline: (i) reduced input resolution and (ii) additive Gaussian noise. For resolution experiments, images were resized to lower resolutions (e.g., $160 \times 160$) and then resized back for model input to simulate loss of detail. For noise experiments, Gaussian noise with standard deviation $\sigma$ was added to normalized images prior to inference.

\section{Results and Analysis}
\subsection{Classification Performance}
The above table summarizes the prototype-level performance for three models: CNN-only (MobileNetV2 classifier), radiomics-only (RandomForest on 8 features), and fusion (RandomForest on concatenated CNN embeddings and radiomics features). These results are reported directly from the implemented notebook experiments and should be interpreted as a proof-of-concept rather than a clinically validated benchmark.

\begin{table}[t]
\caption{Validation performance (four-class MRI classification).}
\label{tab:val_results}
\centering
\begin{tabular}{lcc}
\hline
\textbf{Model} & \textbf{Val. Acc.} & \textbf{Val. Macro-F1} \\
\hline
CNN-only (MobileNetV2) & 0.85 & 0.81 \\
Radiomics-only (RF) & 0.742 & 0.746 \\
Fusion (CNN+Radiomics, RF) & 0.951 & 0.951 \\
\hline
\end{tabular}
\end{table}

\begin{table}[t]
\caption{Test performance (four-class MRI classification).}
\label{tab:test_results}
\centering
\begin{tabular}{lcc}
\hline
\textbf{Model} & \textbf{Test Acc.} & \textbf{Test Macro-F1} \\
\hline
CNN-only (MobileNetV2) & 0.746 & 0.71 \\
Radiomics-only (RF) & 0.670 & 0.631 \\
Fusion (CNN+Radiomics, RF) & 0.751 & 0.724 \\
\hline
\end{tabular}
\end{table}

Fusion improved validation performance significanlty compared with either single branch, suggesting complementary information between the deep embeddings and radiomics descriptors. On the held-out test set, fusion achieved accuracy comparable to the CNN baseline and improved macro-F1 relative to radiomics-only. A notable gap between validation and test performance was observed across models, consistent with dataset shift between training/validation and test splits in this dataset configuration. This gap was treated as a realistic indicator of generalization challenges rather than being optimized away in the prototype.

\begin{figure}[t]
\centering
\includegraphics[width=0.85\linewidth]{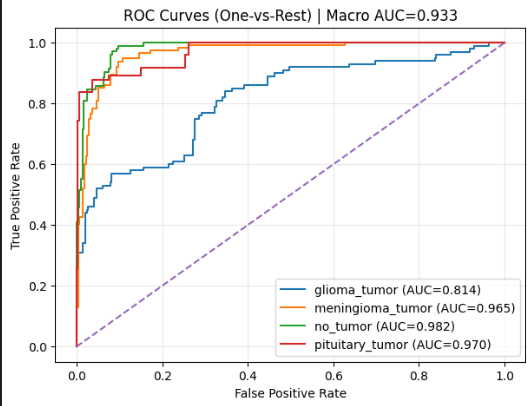}
\caption{One-vs-rest ROC curves for the CNN baseline on the test set. High AUC values for most classes indicate strong ranking performance, while lower AUC for glioma reflects increased class overlap and dataset variability.}
\label{fig:roc}
\end{figure}

To further assess discriminative ability independent of a fixed decision threshold, receiver operating characteristic (ROC) curves were computed for the CNN baseline using a one-vs-rest strategy. Fig.~\ref{fig:roc} reports class-wise ROC curves along with the macro-averaged AUC.

\subsection{Per-Class Behavior}
\begin{figure}[t]
\centering
\includegraphics[width=0.85\linewidth]{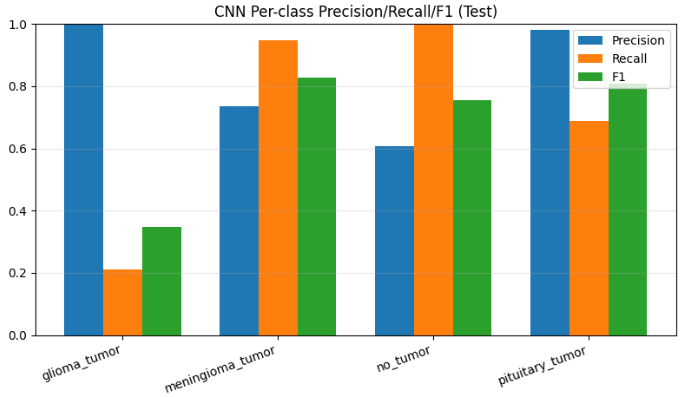}
\caption{Per-class precision, recall, and F1 scores for the CNN baseline on the test set. Glioma exhibits high precision but low recall, indicating conservative predictions, while no-tumor shows high recall, favoring sensitivity in screening-oriented decision support.}
\label{fig:cnn_perclass}
\end{figure}

To better understand class-wise behavior beyond aggregate accuracy and macro-F1 scores, per-class precision, recall, and F1 scores were computed for the CNN baseline. Fig.~\ref{fig:cnn_perclass} highlights asymmetric error patterns across tumor categories.

The CNN model demonstrates strong recall for the \textit{no tumor} class, reducing the risk of false negatives, which is desirable in screening scenarios. In contrast, \textit{glioma} shows very low recall despite high precision, indicating that while predicted glioma cases are reliable, many true glioma instances are missed. This conservative behavior is consistent with inter-class visual overlap and the use of 2D MRI slices without sequence-specific metadata. Performance for \textit{meningioma} and \textit{pituitary} is more balanced, with comparatively higher F1 scores.

\begin{figure}[t]
\centering
\includegraphics[width=0.85\linewidth]{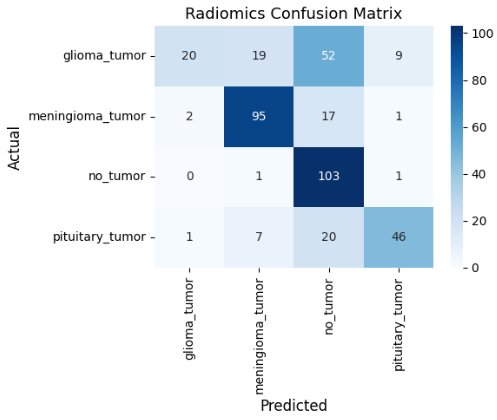}
\includegraphics[width=0.85\linewidth]{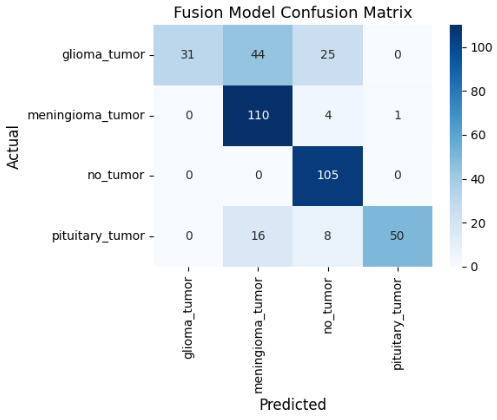}
\caption{Confusion matrices on the test set for the radiomics-only model (top) and the CNN--radiomics fusion model (bottom). Fusion reduces misclassification between tumor and non-tumor cases and improves overall class separability, while residual confusion persists between certain tumor subtypes.}
\label{fig:confusion}
\end{figure}

Classification performance varied across tumor categories, as illustrated by the confusion matrices in Fig.~\ref{fig:confusion}. The radiomics-only model showcases strong performance for the \textit{no tumor} and \textit{meningioma} classes, but shows substantial confusion for \textit{glioma}, with many glioma cases misclassified as non-tumor. This highlights the limitations of simplified handcrafted features in capturing infiltrative or heterogeneous tumor characteristics from 2D MRI slices.

The fusion model significanlty mitigates these failure modes by integrating the CNN-derived embeddings. In particular, fusion reduces tumor versus non-tumor confusion and improves recall for \textit{glioma}, demonstrating that deep features provide complementary spatial and semantic information beyond radiomics alone. Unfortunately, residual confusion remains between certain tumor subtypes, most notably between \textit{glioma} and \textit{meningioma}, which is clinically plausible given overlapping visual characteristics and the absence of some sequence-specific MRI metadata in the dataset. These results reflect some of the realistic challenges associated with dataset variability and low-resource imaging conditions.

\subsection{Robustness Under Degraded Conditions}
Robustness experiments were conducted to simulate low-resource imaging conditions. For resolution degradation, CNN accuracy decreased as resolution was reduced: approximately 0.74 at $224 \times 224$, 0.68 at $160 \times 160$, 0.60 at $112 \times 112$, and 0.48 at $80 \times 80$ in the reported run. For Gaussian noise, CNN accuracy dropped sharply from approximately 0.75 at $\sigma=0.0$ to approximately 0.27 at $\sigma=0.03$ and remained near 0.26--0.27 for higher noise levels up to $\sigma=0.1$ in the reported run. These results highlight a key  concern: CNN-based models may be highly sensitive to noise and reduced resolution, motivating further robustness training and evaluation.

\begin{figure}[t]
\centering
\includegraphics[width=0.75\linewidth]{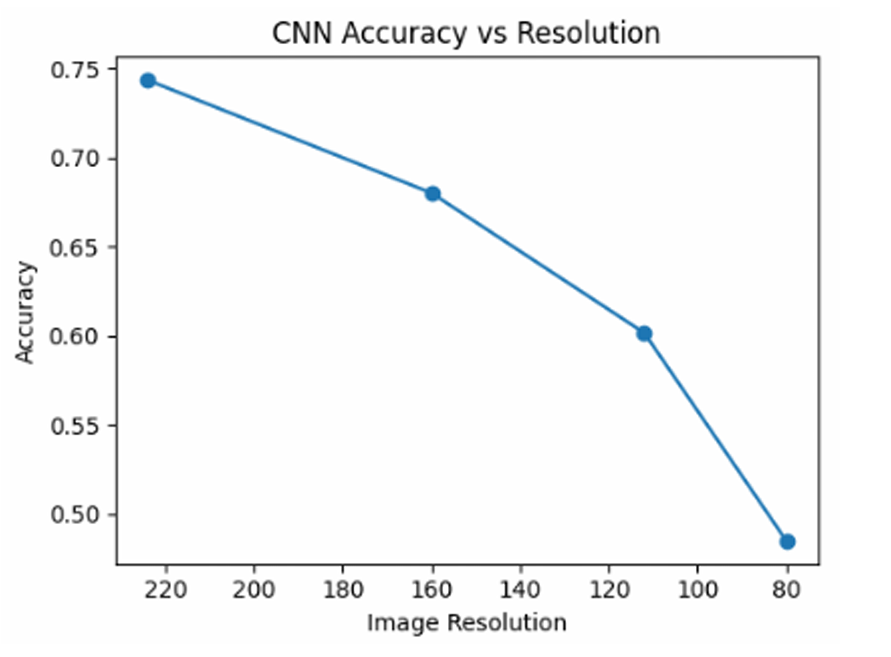}
\includegraphics[width=0.69\linewidth]{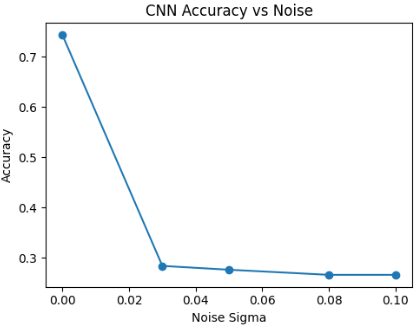}
\caption{Robustness evaluation of the CNN baseline under reduced resolution (top) and additive Gaussian noise (bottom).}
\label{fig:robustness}
\end{figure}

\section{Clinical Relevance and Interpretability}
The proposed system is intended as decision support and as a conceptual exploration of ``virtual biopsy'' workflows. The term ``virtual biopsy'' is used to describe non-invasive inference of tumor characteristics from imaging that may provide complementary information to clinicians, particularly when invasive biopsy is infeasible. The prototype does not replace histopathology, and no diagnostic guarantees are claimed.

Interpretability was addressed at two levels. First, Grad-CAM heatmaps provided qualitative evidence that CNN predictions were influenced by lesion-relevant regions in example cases, enabling plausibility checks by a human reviewer. Second, radiomics feature importance plots provided a numerical summary of which interpretable features contributed to RandomForest decisions. Together, these outputs were designed to support transparent model behavior inspection and to reduce the ``black-box'' perception associated with deep models.

A case-study style visualization was constructed in the slides and notebook outputs, combining the original MRI, Grad-CAM overlay, an automatically derived radiomics mask, and a per-image radiomics profile. Such multimodal explanation is aligned with the broader goal of clinician-facing decision support, where both visual attention and quantitative descriptors can be reviewed.

\section{Limitations and Ethical Considerations}
Following limitations apply. First, the dataset used is a 2D MRI image classification dataset without explicit acquisition metadata; therefore, conclusions about specific MRI sequences (T1, T2, FLAIR) or volumetric tumor characteristics cannot be drawn. Second, the radiomics implementation is a simplified, heuristic feature extraction pipeline that uses Otsu thresholding for masking; this may fail on low-contrast scans and is not equivalent to clinically validated segmentation-based radiomics workflows. Third, the observed validation--test performance gap suggests dataset shift effects that would require broader multi-institutional evaluation and careful domain generalization strategies to mitigate.

From an ethical and safety perspective, using AI models in healthcare should be strongly evaluated for bias, robustness, along with failure modes. This prototype is still a research and educational system for exploring hybrid modeling and explainability under low-resource constraints. Clinical deployment would require much more rigorous validation, regulations, and alignment with institutional policies for patient safety, privacy, and accountability.

\section{Future Work}
Future work will focus on improving clinical realism and robustness. Volumetric (3D) modeling could further better capture tumor context across slices, potentially using efficient 3D CNNs or transformer-based representations when resources permit. Integrating a dedicated segmentation model (e.g., U-Net) could yield more accurate tumor masks, enabling more reliable radiomics and more precise interpretability. Robustness could be improved through augmentation, noise-aware training, domain adaptation, or self-supervised pretraining on diverse MRI sources. Computational efficiency and deployability could be further enhanced using model pruning, quantization, or optimized inference formats (like ONNX). Finally, evaluation on multiple datasets and clinician supported studies would be necessary to assess clinical utility and trust.

\section{Conclusion}
A prototype AI-enhanced ``virtual biopsy'' pipeline was presented for four-class brain tumor MRI image classification, designed with low-resource deployment considerations. The method combined a lightweight MobileNetV2 CNN with interpretable radiomics-style features and a hybrid fusion classifier, and incorporated explainability via Grad-CAM and feature importance. Prototype-level results indicated strong validation performance for the fusion model and highlighted generalization challenges under test distribution shift. Robustness experiments demonstrated significant sensitivity of CNN inference to reduced resolution and noise, which helps in reinforcing the importance of robustness oriented evaluation for low-resource settings. This work is positioned as a non-invasive decision support concept rather than a complete replacement for biopsy, with explicit limitations to support responsible medical AI framing.


\begin{thebibliography}{00}

\bibitem{b1}
A. Sailunaz, M. A. Shovon, and N. M. K. Alam, ``A brain tumor segmentation enhancement in MRI images using U-Net and transfer learning,'' in \textit{Proc. IEEE Int. Conf. Innovations in Information Technology}, 2021.

\bibitem{b2}
A. Rehman, K. Maqsood, and M. A. Khan, ``VGG-SCNet: A VGG Net based deep learning framework for brain tumor detection on MRI images,'' in \textit{Proc. Int. Conf. Computational Intelligence}, 2022.

\bibitem{b3}
S. Pati \textit{et al.}, ``Federated learning enables big data for rare cancer boundary detection,'' \textit{Nature Communications}, vol. 13, no. 1, pp. 1--13, 2022.

\bibitem{b4}
B. Taha and K. Pandey, ``A novel approach for brain tumor classification using bilateral filtering and cascade RF-SVM,'' \textit{Journal of Clinical Neuroscience}, vol. 90, pp. 206--211, 2021.

\bibitem{b5}
L. Mayeta-Revilla \textit{et al.}, ``Radiomics-driven neuro-fuzzy framework for explainability in MRI-based brain tumor segmentation,'' \textit{Frontiers in Neuroinformatics}, vol. 19, 2025.

\bibitem{b6}
N. Siddique, S. Paheding, C. P. Elkin, and V. Devabhaktuni, ``U-Net and its variants for medical image segmentation: A review of theory and applications,'' \textit{IEEE Access}, vol. 9, pp. 82031--82057, 2021.

\bibitem{b7}
L. Wen, X. Li, X. Li, and L. Gao, ``A new transfer learning method based on VGG-19 network for fault diagnosis,'' in \textit{Proc. IEEE 23rd Int. Conf. Computer Supported Cooperative Work in Design (CSCWD)}, 2019, pp. 205--209.

\bibitem{b8}
S. Ghosh, A. Chaki, and K. Santosh, ``Improved U-Net architecture with VGG-16 for brain tumor segmentation,'' \textit{Physical and Engineering Sciences in Medicine}, vol. 44, no. 3, pp. 703--712, 2021.

\bibitem{b9}
N. Sharma, S. Gupta, D. Koundal, S. Alyami, H. Alshahrani, Y. Asiri, and A. Shaikh, ``U-Net model with transfer learning as a backbone for segmentation of the gastrointestinal tract,'' \textit{Bioengineering}, vol. 10, no. 1, p. 119, 2023 \cite{b7,b8}.

\bibitem{b10}
N. Pedano, A. E. Flanders, L. Scarpace, T. Mikkelsen, J. M. Eschbacher, B. Hermes, V. Sisneros, J. Barnholtz-Sloan, and Q. T. Ostrom, ``The Cancer Genome Atlas low-grade glioma (TCGA-LGG) collection,'' 2016.

\bibitem{b11}
K. Clark \textit{et al.}, ``The Cancer Imaging Archive (TCIA): Maintaining and operating a public information repository,'' \textit{Journal of Digital Imaging}, vol. 26, no. 6, pp. 1045--1057, 2013.

\bibitem{b12}
A. Verma, S. N. Shivhare, S. P. Singh, N. Kumar, and A. Nayyar, ``Comprehensive review on MRI-based brain tumor segmentation: A comparative study from 2017 onwards,'' \textit{Archives of Computational Methods in Engineering}, pp. 1--47, 2024.

\bibitem{b13}
C. S. Rao and K. Karunakara, ``A comprehensive review on brain tumor segmentation and classification of MRI images,'' \textit{Multimedia Tools and Applications}, vol. 80, no. 12, pp. 17611--17643, 2021.

\bibitem{b14}
F. Isensee, P. F. J\"ager, S. A. A. Kohl, J. Petersen, and K. H. Maier-Hein, ``Automated design of deep learning methods for biomedical image segmentation,'' \textit{arXiv preprint arXiv:1904.08128}, 2019.


\end{thebibliography}
\end{document}